\documentclass[preprint,12pt]{elsarticle}

\usepackage{latexsym}
\usepackage{epsfig}
\usepackage{graphicx}
\usepackage[english]{babel}

\usepackage{amssymb}

\usepackage[lmargin=2cm,rmargin=2cm,tmargin=2.5cm,bmargin=3.5cm]{geometry}

\usepackage{epstopdf}
\journal{Physics Letters B}
\begin{document}

\begin{frontmatter}

\title{Evidence for nuclear gluon shadowing from the ALICE measurements 
of PbPb ultraperipheral exclusive $J/\psi$ production}

\author[label1]{V. Guzey}
\author[label1]{E. Kryshen}
\author[label2]{M. Strikman}
\author[label1]{M. Zhalov}
% M. Strikman$^2$, M. Zhalov$^1$\\
%$^1$ National Research Center ``Kurchatov Institute'',\\
% Petersburg Nuclear Physics Institute (PNPI), Gatchina, 188300, Russia\\
%$^2$ The Pennsylvania State University, University Park, PA 16802, USA
%}

\address[label1]{National Research Center ``Kurchatov Institute'',
Petersburg Nuclear Physics Institute (PNPI), Gatchina, 188300, Russia}

\address[label2]{The Pennsylvania State University, University Park, PA 16802, USA}

%\date{\today}

%\pacs{24.85.+p,25.20.Lj,25.75.-q} 

%\maketitle

\begin{abstract}
We show
that the recent ALICE measurements of 
exclusive $J/\psi$ production in 
ultraperipheral PbPb collisions at 2.76 TeV 
provide the first direct
experimental evidence for the strong nuclear gluon 
shadowing in lead at $x\sim 10^{-3}$. 
The evidence is based on the comparison of the nuclear 
suppression factor $S(x\approx 0.001)=0.61^{+0.05}_{-0.04}$
found in the analysis of 
the coherent $J/\psi$ photoproduction cross sections measured by ALICE  
with  the nuclear gluon shadowing predicted by the global fits of nuclear parton 
distributions
and by the leading twist theory of nuclear shadowing.

\end{abstract}

%vg

%\linenumbers

%\maketitle

\end{frontmatter}

\section{Introduction}

This brief communication aims to extract 
the nuclear suppression of 
coherent $J/\psi$ photoproduction off nuclei from the data obtained by the ALICE collaboration in 
ultra-peripheral PbPb 
collisions at $\sqrt{s_{NN}} =2.76 \,{\rm TeV}$ at the LHC.

Nucleus--nucleus collisions are considered as 
ultraperipheral collisions (UPCs), if the impact 
parameter $|\vec b|$---the distance between 
the centers of the colliding nuclei in the 
transverse plane of the reaction---is larger than the 
sum of the nuclear radii, i.e., $|\vec 
b|>(2 - 3)R_A$, where $R_A$ is the nuclear radius (for a review of the UPC physics 
see, for example, \cite{Baltz:2007kq}). 
Hadron interactions are strongly suppressed in such collisions 
and, thus, experimentally the UPC events are characterized 
by minimal multiplicity. This results in a relative enhancement 
of electromagnetic processes induced by the high 
flux of photons generated by
ultrarelativistic nuclei which 
scales as $Z^2$, where $Z$ is the charge of the nucleus. 
The photon virtuality is small 
and while its transverse momentum is $\sim 1/R_A$, 
its longitudinal momentum is 
proportional to the large Lorentz factor $\gamma_L$ of the ion producing the photon flux. 
Hence, one can apply the method of equivalent photons to
express the cross section of 
$J/\psi$ production in nucleus--nucleus UPCs 
as a product of the photon flux 
emitted by one of 
the colliding 
nuclei and the cross section of $J/\psi$ 
photoproduction on the other nucleus:
\begin{equation}
\frac {d\sigma_{AA\to AAJ/\psi}(y)} {dy}
=N_{\gamma/A}(y)\sigma_{\gamma A\to AJ/\psi}(y)+
N_{\gamma/A}(-y)\sigma_{\gamma A\to AJ/\psi}(-y) \,.
\label{csupc}
\end{equation}
In Eq.~(\ref{csupc}), 
$N_{\gamma/A}(y) \equiv \omega dN_{\gamma/A}(\omega)/d \omega$ is the photon flux;
$y = \ln(2\omega/M_{J/\psi})=\ln(W^{2}_{\gamma p}/(2\gamma_{L}m_{N}M_{J/\psi}))$ is the $J/\psi$ rapidity, 
where $\omega$ is the photon 
energy (in the laboratory frame), $W_{\gamma p}$ is $\gamma p$ center-of-mass energy,
 $M_{J/\psi}$ is the mass of $J/\psi$ and $m_N$ is the nucleon mass.
The presence of two terms in Eq.~(\ref{csupc})  
is due to the symmetry of PbPb collisions: each nucleus can radiate a 
photon as well as serve as a target. 
The photon flux $N_{\gamma/A}$ can be calculated with reasonable accuracy and, 
therefore, the UPCs can be effectively used to study 
the energy behavior of the 
vector meson photoproduction cross section at high energies.

High energy coherent $J/\psi$ photoproduction on 
nuclei is of a particular interest since 
the large $c$-quark mass, $m_c$, provides a hard scale 
$\mu \ge m_c$ justifying the use of the factorization 
theorem of perturbative QCD (pQCD). 
This allowed one to develop several models predicting the cross section 
of $J/\psi$ photoproduction on nuclear targets at high energies. 
Unfortunately, 
until recently, the progress in experimental studies of 
this process was more than modest: 
about two dozens events have been accumulated in recent measurements of 
$J/\psi$ photoproduction in AuAu UPCs at RHIC 
at $\sqrt {s_{NN}}=200$ GeV~\cite{phenix1}.  

In 2011, the ALICE collaboration measured the yield of coherent $J/\psi$ photoproduction 
in PbPb UPCs in the rapidity range of $|y|\le 0.9$  with 
the central barrel~\cite{alice1} and 
in the range of $-3.6\le y \le -2.6$ covered by 
the muon spectrometer~\cite{alice2}.
This allowed one to obtain 
the cross section
$\sigma_{PbPb\rightarrow PbPbJ/\psi }(y)$ at two values of rapidity:
\begin{eqnarray}
\sigma_{PbPb\to PbPbJ/\psi} (y=0) &\approx&
2.38^{+0.34}_{-0.24} {\rm (stat. + syst.)}\ {\rm mb} \, , %. 
\label{csbar}
\\
\sigma_{PbPb\to PbPbJ/\psi}(y=-3.1)
&\approx &
1.00 \pm 0.18 {\rm(stat.)} {}^{+0.23}_{-0.26} {\rm (syst.)}\ {\rm mb} \, .
\label{csspect}
\end{eqnarray}

These values of 
$\sigma_{PbPb\rightarrow PbPbJ/\psi}(y)$ were compared \cite{alice1,alice2}
to a number of predictions and appeared to be in a better agreement
with those which calculated coherent
$J/\psi$ photoproduction on nuclear targets in the leading order (LO) pQCD 
taking into account the nuclear gluon shadowing~\cite{rsz,ba}. 

However, it is reasonable 
to reduce as much as possible the model dependence
in the comparisons of the experimental cross
sections with different model calculations. 
We believe that the best strategy to achieve this goal 
 is to analyze the ALICE results in terms of the nuclear 
suppression factor $S(W_{\gamma p})$.
We define 
$S(W_{\gamma p})$ 
through
the ratio of the  
experimentally 
measured coherent photoproduction cross section at a 
given $W_{\gamma p}$ to the cross section calculated 
in the 
 impulse approximation (IA)
which neglects all nuclear effects except for coherence:
\begin{equation}
S (W_{\gamma p}) \equiv 
%vg {\biggl [
\left[
\frac {\sigma^{\rm exp}_{\gamma Pb\rightarrow J/\psi Pb}
(W_{\gamma p})} 
{\sigma^{\rm IA}_{\gamma Pb\rightarrow J/\psi Pb}(W_{\gamma p})}
%vg \biggr ]}^{1/2} \, .
\right]^{1/2} \,.
\label{ratio}
\end{equation}
Such a definition of $S(W_{\gamma p})$ for coherent vector meson photoproduction 
on nuclear targets corresponds to the standard estimate of nuclear suppression
in terms of $A_{\rm eff}/A$~\cite{Spital:1974cf}.
Since the nucleus remains intact in the considered process, 
the transverse momentum  distribution of $J/\psi$ is dictated by the 
elastic nuclear form factor $F_A (t)$. 
Hence,  
the cross section in the impulse approximation can be  
written as:
\begin{equation}
{\sigma^{\rm IA}_{\gamma Pb\rightarrow J/\psi Pb}(W_{\gamma p})}=
{{d\sigma_{\gamma p\rightarrow J/\psi p }(W_{\gamma p},t=0)\over dt}
\Phi_A(t_{\rm min})} \,.
\label{sigmaIA}
\end{equation}
In Eq.~(\ref{sigmaIA}),
$d\sigma_{\gamma p\rightarrow J/\psi p}(W_{\gamma p},t=0)/dt$ 
is the forward differential cross section of 
$\gamma +p\rightarrow J/\psi+p$ which can be extracted 
from the experimental data~\cite{hera};  
$\Phi_A (t_{\rm min})$ is defined as the integral over the nuclear form factor $F_A(t)$ squared:
\begin{equation}
\Phi_A(t_{\rm min})=\int \limits_{t_{\rm min}}^{\infty} dt 
{\left|F_A (t)
\right|
}^2 \,,
\label{fi}
\end{equation}
where 
$t_{\rm min}=-p_{z,{\rm min}}^2=-[M^2_{J/\psi}/(4\omega \gamma _L)]^2$ is
determined by the minimal longitudinal momentum 
transfer  $p_{z,{\rm min}}$ 
characterizing the coherence length which becomes 
important in the low energy domain. 
In the case of Pb, 
the nuclear form factor
 \begin{equation}
F_A (t)=\int d^2\vec{b}\,dz\, e^{i{\vec p_{t}}\cdot {\vec b}}
e^{i{p_{z}}\cdot z}\rho_A ({\vec b},z) \, ,\qquad
F_A (0)=A \, ,
\label{formfac}
\end{equation} 
can be calculated with a small uncertainty 
since the nuclear density distribution $\rho_A ({\vec r})$ is well 
known from the electron--lead and proton--lead 
elastic scattering experiments~\cite{dens}.

It is important to point out
that the suppression factor $S(W_{\gamma p})$ 
is practically  model independent since 
the estimate of the cross section in the impulse approximation is
based on experimental data.

%\newpage
\section{Calculation of the suppression factor}

\subsection{Cross section of coherent $J/\psi$ 
photoproduction in ALICE measurements}

In this subsection, we determine the coherent $J/\psi$ photoproduction 
cross section $\sigma_{\gamma Pb\rightarrow J/\psi Pb}(W_{\gamma p})$ 
from the values of $\sigma_{PbPb\rightarrow PbPbJ/\psi}(y)$ measured by ALICE. 
In general, the extraction of $\sigma_{\gamma Pb\rightarrow J/\psi Pb}(W_{\gamma p})$ is not 
straightforward 
due to the presence of two terms in Eq.~(\ref{csupc}).
However, this problem is not present
at $y=0$ since the energies of the photons emitted by both colliding ions
%vg ($\omega =(M_{J/\psi}/2)e^{y}$ and $\omega =(M_{J/\psi}/2)e^{-y}$) 
are equal 
in this case. Thus, we obtain:
\begin{equation}
\sigma_{\gamma Pb\to J/\psi Pb}(W_{\gamma p} = 92.4\,{\rm GeV})=\frac {\sigma_{PbPb\to PbPbJ/\psi}(y=0)} 
{ 2N_{\gamma/Pb}(y=0)} \,.
\label{csbar1}
\end{equation}

On the other hand, at forward  rapidity $y=-3.1$, there are low-energy and high-energy 
contributions. 
Calculations predict the strong dominance of the low-energy contribution (more than $95\%$)
in Eq.~(\ref{csupc}) due 
to a steep falloff of the photon flux for high photon energies. Therefore, 
in this case we obtain: 
\begin{equation}
\sigma_{\gamma Pb\to J/\psi Pb}(W_{\gamma p} = 19.6\,{\rm GeV})=\frac {\sigma_{PbPb\to PbPbJ/\psi}(y=-3.1)}  
{N_{\gamma/Pb}(y=-3.1)} \,,
\label{csspect1}
\end{equation}
which is valid with a 5\% error 
that is well below the experimental uncertainties.

Next we need
to calculate the photon flux $N_{\gamma /Pb}(y)$ 
produced by Pb nuclei with the energy of 1.38 $A$TeV. 
Experimentally, coherent $J/\psi$ events are selected by requiring 
only two daughter leptons and the otherwise empty detector.
This means that the strong interactions between 
colliding nuclei should be suppressed and, hence,
the impact parameter is 
required to be larger than the sum of the nuclear radii: $b>2R_A$. 
Actually, the sharp cutoff at $2R_A$ can be 
improved by using a more accurate 
approximation in the
calculation of the strong interaction suppression. Indeed,
using the profile factor 
\begin{equation}
 \Gamma_{AA}({\vec b})=
 \exp\biggl (-\sigma_{NN}
 \int \limits^{\infty}_{-\infty}dz\int d^2\vec{b}_1\,
 \rho_A(z,{\vec b_1})\rho_A(z,{\vec b}-{\vec b_1})\biggr ) \,,
\label{eq:gamma_AA}
\end{equation}
where $\sigma_{NN} = 80$ mb is the total nucleon--nucleon ($NN$) cross section 
at $\sqrt{s} = 2.76$ TeV~\cite{Pdg2012},
the photon flux can be calculated as the following 
convolution: 
\begin{equation}
 N_{\gamma /A}(\omega )= \int \limits_{2 R_A}^{\infty} d^2\vec{b}\, 
\Gamma_{AA}({\vec b}) 
 N_{\gamma /A}(\omega ,\vec b) \,.
 \label{flux}
\end{equation}
The photon flux at the 
transverse distance (impact parameter) $\vec b$ from the 
center of a fast moving heavy nucleus reads, see, e.g., \cite{vidovic}:
\begin{equation}
N_{\gamma /A}(\omega,\vec b )=\frac{Z^2\alpha}{\pi^2}
\left|
\int_0^\infty dk_\bot
{{k^{2}_\bot F_A (k^{2}_{\bot}+{{\omega ^2}/{\gamma ^{2}_{L}}})} 
\over {{k^{2}_{\bot}+{{\omega ^2}/{\gamma ^{2}_{L}}}}}}\,
J_{1}(bk_{\bot})
\right|^2 \,,
\label{bflux} 
\end{equation}
where $\alpha$ is the fine-structure constant; $J_{1}$ is the Bessel function of the first kind.
 
Using Eqs.~(\ref{eq:gamma_AA})--(\ref{bflux}), we obtain the following values of the photon flux:
\begin{eqnarray}
 N_{\gamma/Pb}(y=0) &=&  67.7 \pm 3.4 \,, \nonumber\\ 
 N_{\gamma/Pb}(y=-3.1) &=& 163.9 \pm 8.2 \,.
 \label{eq:flux}
\end{eqnarray}
The quoted uncertainties were estimated by utilizing different nuclear density distributions 
in the calculation of the nuclear form factor and the profile factor.

Using Eqs.~(\ref{csbar1}) and (\ref{csspect1}), 
the values of the cross sections measured by ALICE [Eqs.~(\ref{csbar}) and (\ref{csspect})] and our estimates of 
the photon flux [Eq.~(\ref{eq:flux})],  we obtain
the following values of the $J/\psi$ photoproduction cross section:
%vg \approx -> =
\begin{eqnarray}
\sigma_{\gamma Pb\to J/\psi Pb}(W_{\gamma p} = 92.4\,{\rm GeV})
&=& 17.6^{+2.7}_{-2.0}\ {\rm {\mu}b} \,, \nonumber\\
\sigma_{\gamma Pb\to J/\psi Pb}(W_{\gamma p} = 19.6\,{\rm GeV})
&=& 6.1^{+1.8}_{-2.0} \ {\rm {\mu}b} \,,
\label{sigmaEXP}
\end{eqnarray}
where the experimental errors and the flux uncertainty were added in quadrature.

\subsection{$J/\psi$ photoproduction on Pb in the impulse approximation}

To determine the nuclear suppression factor $S(W_{\gamma p})$
from Eq.~(\ref{ratio}), we 
need to calculate the coherent $J/\psi$ photoproduction 
cross section in the impulse approximation,  
$\sigma^{\rm IA}_{\gamma Pb\rightarrow J/\psi Pb}(W_{\gamma p})$, 
which is given by Eq.~(\ref{sigmaIA}).

\begin{figure}[htb]
\centering
\epsfig{file=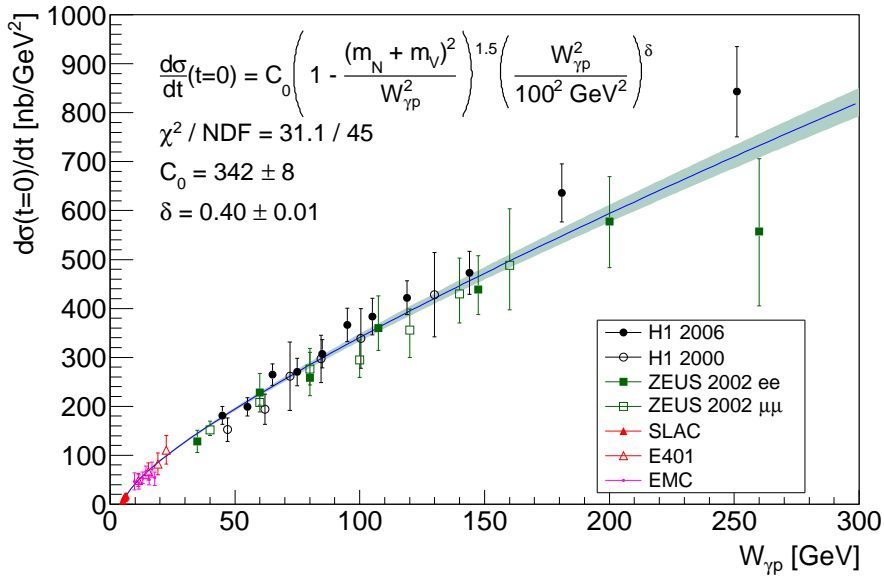, width=12cm}
\caption{The fit to the forward $J/\psi$ photoproduction cross section data~\cite{hera}.}
\label{forcs}
\end{figure}

The forward differential $\gamma +p\rightarrow J/\psi+p$ cross section at 
$W_{\gamma p}=19.6\ {\rm GeV}$ and $W_{\gamma p}=92.4\ {\rm GeV}$ can be 
extracted from HERA, FNAL and CERN measurements~\cite{hera}\footnote{
Exclusive $J/\psi$ production has been measured at $y=0$ in 
ultraperipheral $p\bar p$ collisions by the CDF collaboration at 
Tevatron~\cite{jpsitevatron}. The ${\gamma p\rightarrow J/\psi p}$ cross section
obtained from their value of $\sigma_{p\bar p \rightarrow p\bar pJ/\psi}(y=0)$ 
is in reasonable agreement with the HERA measurements.
Recently, the LHCb collaboration measured the yield of $J/\psi$ 
at the forward rapidities of $2<y<4.5$ in proton--proton UPCs at 7 TeV~\cite{lhcb}. 
This data allowed one to extract the 
$\gamma p\to J/\psi p $ cross section
at the $\gamma p$ center-of-mass energies of 
$0.6\,{\rm TeV}<W_{\gamma p}< 1\,{\rm TeV}$.}. 
 A compilation of the experimental results is shown in Fig.~\ref{forcs}. 
The $10\ {\rm GeV}<W_{\gamma p}<25\ {\rm GeV}$ range of energies 
corresponding to the ALICE muon spectrometer acceptance in the measurement 
of $J/\psi$ production in PbPb UPCs at 2.76 TeV was studied in the old 
proton-target experiments at FNAL and CERN. Statistics in those 
experiments was very low resulting in large experimental errors. 
The forward $J/\psi$ photoproduction cross section at higher energies was 
measured by the H1 and ZEUS collaborations at HERA. 
As can be seen in Fig.~\ref{forcs}, the cross sections measured by these two experiments do not agree 
well, with the most recent H1 measurement being systematically higher over 
the entire energy range. 

The data in Fig.~\ref{forcs} was fitted using the following pQCD motivated 
expression~\cite{Strikman:2005ze}:
\begin{equation}
\frac {d\sigma_{\gamma p\rightarrow J/\psi p}(W_{\gamma p},t=0)} {dt}=
C_0\biggr [1-\frac {(M_{J/\psi}+m_N)^2} {W^2_{\gamma p}}\biggl ]^{1.5}
\biggr [\frac {W^2_{\gamma p}} {100^2\ {\rm GeV^2}}\biggl ]^{\delta} \,,
\label{elcs}
\end{equation}
The values of the free parameters $C_0$ and $\delta$ were determined from the fit, resulting in  
$C_0 =342 \pm 8\ {\rm nb/GeV^2}$ and $\delta =0.40 \pm 0.01$. 
Then, 
the corresponding values of the forward cross section are:
\begin{eqnarray}
\frac {d\sigma_{\gamma p\rightarrow J/\psi p}(19.6\ {\rm GeV}, t=0)}{dt} 
&=& 86.9 \pm 1.8\ {\rm nb/GeV^2} \,, \nonumber \\
\frac {d\sigma_{\gamma p\rightarrow J/\psi p}(92.4\ {\rm GeV}, t=0)}{dt} 
&=& 319.8 \pm 7.1\ {\rm nb/GeV^2} \,.
\label{dsdt}
\end{eqnarray}

To calculate $\Phi_A(t_{\rm min})$ and to estimate its uncertainty,
we evaluate $\Phi_A(t_{\rm min})$ using three different nuclear form factors. 
In particular, we used the
 analytic parametrization of $F_{\rm Pb} (t)$
from StarLight~\cite{starlight}, which
is widely used in analyses of experimental data as a UPC generator. 
We also calculated the nuclear form factor using the Hartree--Fock--Skyrme nuclear density 
distribution $\rho_{\rm Pb} (\vec r )$  
and the Woods--Saxon density distribution with the parameters 
$R_{\rm Pb} = 6.62 \pm 0.06$ fm and $a = 0.546 \pm 0.01$ fm~\cite{Vries1987}. 
These distributions provide the good description of the elastic electron--lead 
and proton--lead scattering data.
The three estimates agree within 5\% 
error
for both considered energies. In the 
following calculations, we use the $\Phi_A(t_{\rm min})$ values corresponding to 
the Woods--Saxon density and assign a $\pm 5\%$ uncertainty:
\begin{eqnarray}
\Phi_A(t_{\rm min}(W_{\gamma p}=19.6\ {\rm GeV})) &=& 127.2\pm 6.4 {\rm\ GeV^2} \,, \nonumber \\
\Phi_A(t_{\rm min}(W_{\gamma p}=92.4\ {\rm GeV})) &=& 149.2\pm 7.5 {\rm\ GeV^2} \,.
\label{PhiWS}
\end{eqnarray}
Combining Eqs.~(\ref{sigmaIA}), (\ref{dsdt}) and (\ref{PhiWS}),
we obtain the 
following values of the $J/\psi$ photoproduction 
cross sections in the impulse approximation:
\begin{eqnarray}
\sigma^{\rm IA}_{\gamma Pb\to J/\psi Pb}(W_{\gamma p} = 19.6\,{\rm GeV})
&=& 11.1 \pm 0.6\ {\rm {\mu}b} \,, \nonumber \\
\sigma^{\rm IA}_{\gamma Pb\to J/\psi Pb}(W_{\gamma p} = 92.4\,{\rm GeV})
&=& 47.7 \pm 2.6\ {\rm {\mu}b} \,,
\label{ia}
\end{eqnarray}
where the uncertainty of the fit to the forward $J/\psi$ cross section and 
the $\Phi_A(t_{\rm min})$ uncertainty were added in quadrature.

\subsection{Estimation of the suppression factor $S(W_{\gamma p})$}

Combining Eqs.~(\ref{ratio}), (\ref{sigmaEXP}) and (\ref{ia}), 
 we estimate the values of the nuclear suppression factor 
$S(W_{\gamma p})$ at the $W_{\gamma p}$ values corresponding to the ALICE 
measurements~\cite{alice1,alice2}:
\begin{eqnarray}
 S(W_{\gamma p} &=& 19.6\,{\rm GeV}) = 0.74^{+0.11}_{-0.12} \,, \\
 \label{supfac1}
S(W_{\gamma p} &=& 92.4\,{\rm GeV}) = 0.61^{+0.05}_{-0.04} \,.
\label{supfac2}
\end{eqnarray}
The dominant part of uncertainties in our estimates comes from the large 
experimental errors in the measured cross section of $J/\psi$ production 
in PbPb UPCs.

\section{Comparison of the suppression factor with theoretical predictions}

The suppression factor $S(W_{\gamma p})$ 
that we extracted from the ALICE data 
can be compared to theoretical predictions for $S(W_{\gamma p})$ made in different 
models and approaches.

The earliest approach to the calculation of the cross 
section of vector meson photoproduction on nuclei is 
based on the vector meson dominance (VMD) model, see the review 
in~\cite{Bauer:1977iq}. In this approach, 
a high energy photon  
converts into a vector meson at a long distance 
(time) before the target. Then, the vector meson interacts coherently
with the nucleus by means of 
multiple interactions with the target nucleons. This process
of hadron-nucleus interaction is 
usually described by the Glauber theory. In assumption that the 
multiple interactions don't distort the shape of the transverse momentum
distribution of the vector meson but result only in the absorption effects
the suppression factor for coherent vector meson production 
on the nucleus can be 
estimated as
\begin{equation}
S_A(W_{\gamma p})= 
%\left[
\frac {\sigma_{VA}(W_{\gamma p})}
{A\sigma_{VN}(W_{\gamma p})}
%\right]^2 
\,,
\label{slsup}
\end{equation}
$\sigma_{VN}$ is the total vector meson--nucleon cross section.
The total vector meson--nucleus cross section $\sigma_{VA}$ can be calculated 
in the optical limit of the Glauber theory:
\begin{equation}
\sigma_{VA}(W_{\gamma p})=2 \int d^2 {\vec b}
\left[1-\exp\left\{-\frac{\sigma_{VN}(W_{\gamma p})}{2}T_A(\vec b)\right\}\right] \,,
\label{gltsig}
\end{equation}
where $T_A(\vec b)=\int \limits_{-\infty}^{\infty} \rho_{A} ({\vec b},z)dz$.

In the case of $J/\psi$ photoproduction, the application of the VMD approach is 
problematic because the  standard VMD model does not take into account 
that the space--time picture of production of heavy onium states 
involves three stages: 
the production of  ``frozen'' small-size $q\bar q$ configurations, their  
interaction with the  target, and the conversion of $q\bar q$ into the 
final-state onium.
This space--time picture can be modeled using the color dipole framework
to calculate the cross section of the interaction of the $c\bar c$ 
configuration with the nucleon 
and the Glauber theory (eikonal model) to describe the propagation 
of the dipoles of the fixed transverse size 
through the nuclear medium. 
We estimated the nuclear suppression factor 
using the phenomenological Golec-Biernat--Wusthoff dipole 
cross section~\cite{GolecBiernat:1998js}:
\begin{equation}
\sigma_{VN}(W_{\gamma p})=\sigma_{c {\bar c}N}(W_{\gamma p})=\sigma_{0}\biggl [1-\exp
\left( - 0.25{\langle d \rangle ^2} \left(\frac {x_0} {x}\right)^{2\lambda}\right) \biggr ] \,,
\label{gbw}
\end{equation}
where $\langle d \rangle \approx 0.25$ fm is 
the average size of the $c\bar c$ configuration in $J/\psi$;  
$\sigma_0 =29.12$ mb, $\lambda=0.277$ and $x_0 =0.000041$
were obtained in \cite{GolecBiernat:1998js} from the fit to the nucleon DIS data at small
$Q^2$ and $x$. One can see from Eq.~(\ref{gbw}) that
this model assumes a gradual increase of the cross section with a decrease of $x$ and, hence,
an eventual onset of the saturation regime at very small $x$.

\begin{figure}[htb]
\centering
\epsfig{file=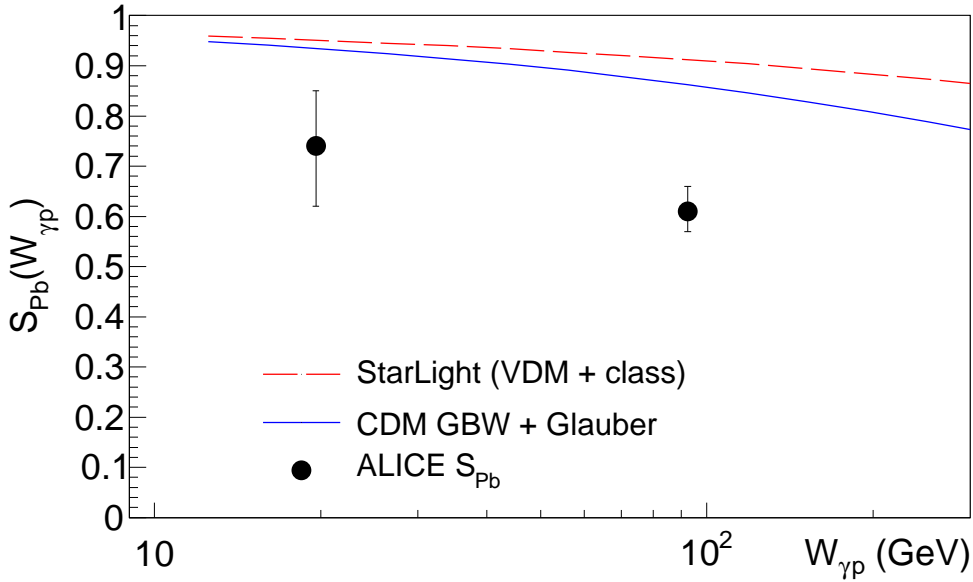, width=13 cm}
\caption{Comparison of the ALICE suppression factors with the estimates in the Glauber model 
with the color dipole cross section and in the Starlight approach.}
\label{shadgl}
\end{figure}

Figure~\ref{shadgl} presents 
the nuclear suppression factor for lead, 
$S_{\rm Pb}(W_{\gamma p})$, as a function of $W_{\gamma p}$.
The result of the calculation using Eqs.~(\ref{slsup})--(\ref{gbw})
is shown by the blue solid line. One can see from the figure that the predicted 
nuclear suppression  
is too small compared to the ALICE results, which are shown as two points with the 
corresponding 
error bars.

It is worth noting here that for the discussed kinematics, 
the results for the dipole--nucleon cross section 
obtained in different dipole models are rather close 
since they are constrained well by the DIS data for these energies.
Note also that in the discussed model, the nuclear shadowing  effect is 
driven by the 
$\sigma_{c\bar c N}$ dipole cross section and, hence, shadowing is suppressed 
(a higher twist effect)  
for the dipoles of such a small size.

One can also estimate the nuclear suppression  in the approach
used in StarLight generator of ultraperipheral collisions~\cite{starlight}. 
In StarLight, the total  cross section
$\sigma_{J/\psi A}(W_{\gamma p})$ is calculated  
using
the classical probabilistic formula\footnote{Note that 
the classical probabilistic formula
and the Glauber formula give close values of the total $VA$ cross section only when 
$\sigma_{VN}T_A(\vec b)\ll  1$.}:
\begin{equation}
\sigma_{VA}(W_{\gamma p})=\int d^2{\vec b}
\biggl [1-\exp\biggl \{-{{\sigma_{VN}(W_{\gamma p})}}T_A(\vec b)\biggr \}\biggr ] \,,
\label{gltsig_2}
\end{equation}
while the values of  $\sigma_{J/\psi N}(W_{\gamma p})$ are  
found 
from the HERA
experimental data on $\gamma p\rightarrow J/\psi p$ cross 
sections~\cite{hera} using the VMD model. 
As a result, the estimated value of $\sigma_{VN}$ is rather small, which leads to
the small nuclear suppression factor $S_A(W_{\gamma p})$. 
Indeed, the prediction for $S_{\rm Pb}(W_{\gamma p})$ calculated 
using Eqs.~(\ref{slsup}) and (\ref{gltsig_2})
with the parameters from StarLight\cite{starlight}, which 
is shown by the red dashed line in Fig.~\ref{shadgl}, significantly overestimates 
the data points that we model-independently extracted from the ALICE data.

It is of particular interest to compare the nuclear suppression 
found from the analysis of the ALICE data to the corresponding predictions of
perturbative QCD. 
At high energies and small transverse momenta  
of $J/\psi$ ($W_{\gamma p} \gg M_{J/\psi} \gg p_t$), in the 
leading order pQCD,  the cross section of 
coherent $J/\psi$ photoproduction on the proton is 
proportional to the proton gluon density $G_p (x,\mu^2)$ 
squared~\cite{Ryskin:1992ui,Brodsky:1994kf}:
\begin{equation}
\frac {d\sigma_{\gamma p\rightarrow J/\psi p }(W_{\gamma p},t=0)} {dt}= 
C({\mu}^2)
\biggl [xG_p (x,\mu^2 )\biggr ]^2 \,,
\label{dcslo}
\end{equation} 
where $x=M^2_{J/\psi}/W^2_{\gamma p}$  is the fraction of the proton 
plus-momentum carried by the gluons; $\mu^2$ is the hard scale.
In the approximation that the Fermi motion of the quarks in charmonium is neglected,
   the prefactor 
$C(\mu ^2)=M^3_{J/\psi}\Gamma_{ee} {\pi}^3 {\alpha_{s}}^2(\mu^2 
)/(48{\alpha_{em}}\mu^8)$,
where $\Gamma_{ee}$ is the width of the $J/\psi$ electronic decay.

\begin{figure}[htb]
\centering
\epsfig{file=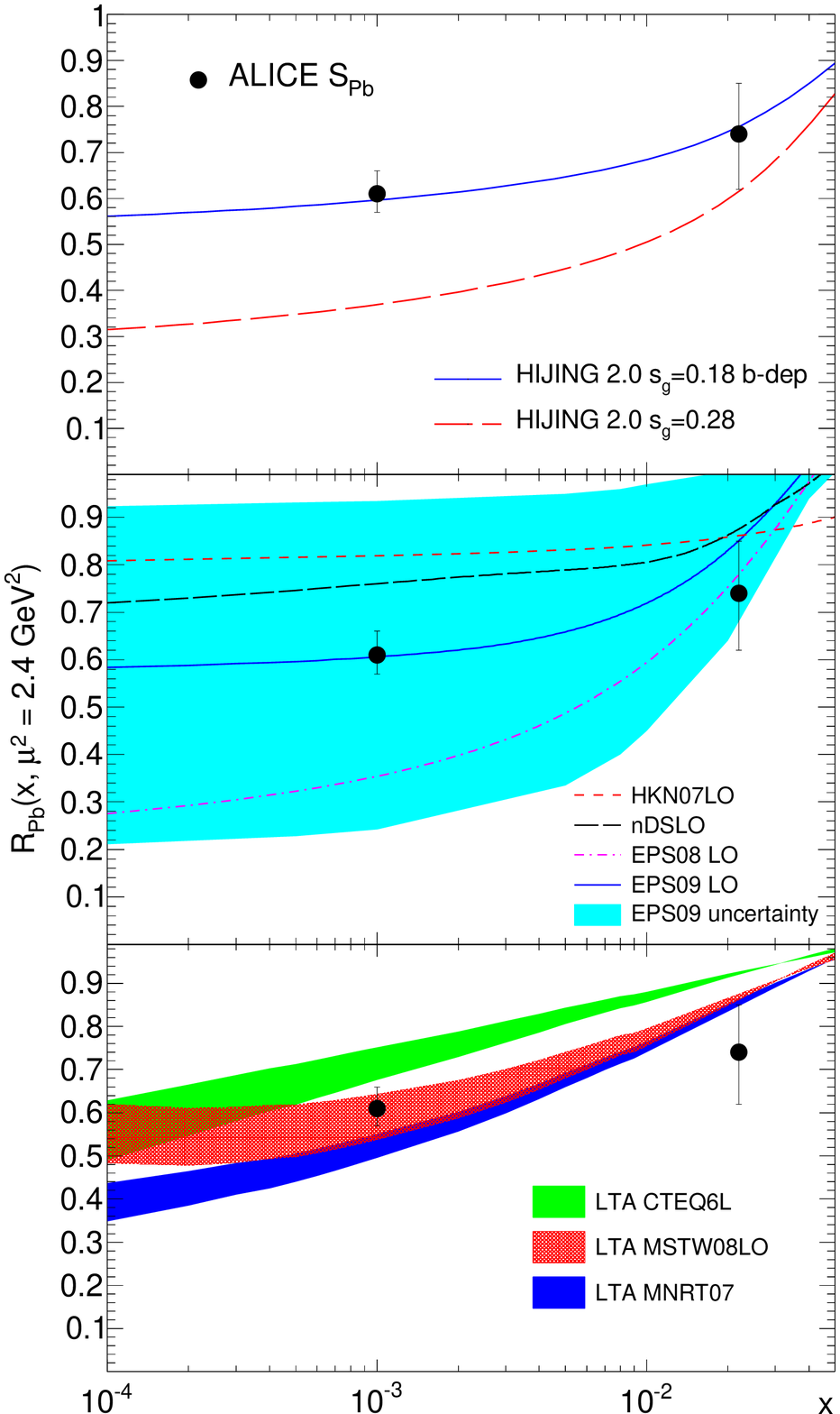, width=12 cm}
\caption{Comparison of the ALICE suppression factors with predictions of the nuclear 
gluon shadowing in HIJING~2.0 (top), global QCD fits (middle), and
in the leading twist approximation (bottom).}
\label{shad}
\end{figure}

It is worth noting
that the accuracy of the LO pQCD calculations
of the $J/\psi$ photoproduction cross section is  still a subject of 
 discussions, see, e.g., \cite{Ryskin:1995hz,hoodbhoy,fks}. 
In particular, the value of the hard scale $\mu^2$ in the the gluon density 
 is not fixed reliably. 
There are also some uncertainties in estimates of
the skewness of the  gluon distributions, relativistic effects
in the charmonium wave function, and higher order corrections. 
Some of the corrections increase 
the cross section,
others -- suppress it. However, there is a general consent that these effects 
mainly influence the absolute value of the cross section but not its energy
dependence. 
The total uncertainty of the LO pQCD 
predictions is estimated in \cite{Ryskin:1995hz,hoodbhoy} to be about
$30\%$ or less,
while \cite{fks} suggests a larger uncertainty.

Extending Eq.~(\ref{dcslo}) for the description of $J/\psi$ 
production on nuclei and accounting for the transverse momentum distribution 
dictated by the nuclear form factor, one can easy find
\begin{equation}
{\sigma^{\rm pQCD}_{\gamma A\rightarrow J/\psi A }(W_{\gamma p})}=
\frac {d\sigma_{\gamma p\rightarrow J/\psi p }(W_{\gamma p},t=0)} {dt} 
\left[\frac{G_A (x,\mu^2 )}{AG_N (x,{\mu}^2)}\right]^2 {\Phi_A(t_{\rm min})} \,.
\label{LTpsi}
\end{equation}

In the impulse approximation, $G_A (x,{\mu}^2)=AG_N (x,{\mu}^2)$ 
and, hence, the nuclear suppression factor for coherent $J/\psi$ photoproduction on nuclei
is
\begin{equation}
S_A(W_{\gamma p})=
%\frac 
%{\sigma^{\rm pQCD}_{\gamma A\rightarrow J/\psi A }(W_{\gamma p})} 
%{\sigma^{\rm pQCD, IA}_{\gamma A\rightarrow J/\psi A }(W_{\gamma p})}
%=  
%\left[
\frac {G_A(x,{\mu}^2)} {AG_N(x,{\mu}^2)} 
%\right]^2 
\equiv
%\left [
R(x,\mu ^2)
%\right ]^2 
\,.
\label{lorat}
\end{equation}

Hence, in the leading order pQCD, the suppression of 
coherent $J/\psi$ photoproduction on nucleus as compared to the impulse approximation
results from the coherent nature of the small $x$ 
screening of the gluon field of  the nucleus
which is generally accepted to be characterized 
by the $R(x,\mu ^2)$ factor\footnote{Note that a consistent 
treatment within the LO pQCD 
requires the use of
 the same proton gluon density $G_p (x,\mu ^2)$ in 
the calculation of the forward $\gamma p\rightarrow  J/\psi p$ cross section 
and in the definition of $R(x,\mu ^2)$.}.

In the top panel of Fig.~\ref{shad}, we compare the values 
of ${S(W_{\gamma p})}$
obtained in our analysis of the ALICE data (two black solid circles)
at $x=0.022$ and $x=0.001$ corresponding to the  energies of 
$W_{\gamma p}=19.6$ GeV and $W_{\gamma p}=92.4$ GeV 
 to the $x$ dependence of the parametrization of the nuclear 
gluon shadowing factors $R(x)$ 
 used in HIJING~2.0 generator~\cite{hijingold, hijingnew}. 
 In this approach, the nuclear gluon shadowing is characterized by  
 the parameter $s_g$ and, contrary to pQCD, it does not depend 
on the scale. In the older 
 version of HIJING~\cite{hijingold}, the values of $s_g$ were chosen to be 
in the range of $0.24-0.28$. The nuclear gluon shadowing in this case---which is  
shown by the red dashed curve---is too strong compared to the ALICE suppression factor
at $x\approx 0.001$.
 More recent versions of HIJING include the impact parameter dependence 
of the nuclear gluon shadowing~\cite{hijingnew} and use the values 
of $s_g= 0.17 - 0.22$ determined from fits to the RHIC hadron production data
within a two-component mini-jet model. The parametrization
with $s_g \approx 0.18$ describes the ALICE values very well, see the blue solid curve in the top
panel of  Fig.~\ref{shad}.

In the middle panel of Fig.~\ref{shad}, we compare the nuclear suppression factor
found from the analysis of the ALICE data to the $x$ 
dependence of the nuclear gluon shadowing
factors obtained using several nuclear parton distribution functions (PDFs). 
These nuclear PDFs are the 
results of the global QCD fits based
on the data on deep inelastic and Drell--Yan processes on
nuclei. In particular, we consider HKN07LO~\cite{hkn}, 
nDSLO \cite{nds}, EPS08LO and EPS09LO~\cite{eps09}.
In accordance with~\cite{Ryskin:1992ui}, we 
take $\mu ^2 =M_{J/\psi}^2/4$, which is close to the
$c$-quark mass squared. Note that a somewhat larger value of $\mu ^2$ is
preferred by the analysis of~\cite{fks}.

From the comparison shown in the middle panel of Fig.~\ref{shad}, we see 
that the HKN07LO, nDSLO and EPS08LO predictions for
$R(x,\mu ^2=2.4 \,\,{\rm GeV}^2)$ are disfavored by the strong contradiction with
the nuclear suppression found by ALICE at $x\approx 0.001$:
while HKN07LO and nDSLO predict too weak shadowing, the 
EPS08 shadowing is 
 too strong.
A good agreement is 
observed for  the central set
of the EPS09LO nuclear gluon shadowing factor 
(blue solid line). However, one has to admit that
the uncertainties of EPS09LO
(turquoise shaded area) are very large. 

It is worth noting here that
the main problem in the determination of the nuclear 
gluon shadowing at $x \sim 10^{-3}$
 in the global QCD fit analyses is the
 lack 
of high quality data sensitive to the nuclear gluon PDFs 
not only at these values of $x$, but also at 
larger $x\approx 0.01$.  
As a result, in these fits, almost any values 
of $ R_A(x\sim 10^{-3}, Q_0^2) $ between 0 and 1 
are allowed leading to strong differences between different analyses. 
Moreover, results of the fits
strongly  depend on the data selected for a given analysis. 
In particular, in addition to the DIS and DY data, the EPS08 analysis
included in their fit the BRAHMS data~\cite{Arsene:2004cn} 
on forward high $p_t$ pion production in dA collisions at RHIC assuming that
the nuclear modification of the pion yield in this kinematics is due to
the gluon shadowing. This resulted in a very strong
effect of nuclear gluon shadowing. However, already in the EPS09 analysis,
this data was excluded from the fit since it was difficult 
to separate the gluon shadowing from 
other pQCD mechanisms also leading to the suppression of 
the yield of forward high $p_t$ pions.
Instead, the data on inclusive neutral pion production in dA collisions
from PHENIX~\cite{Adler:2006wg}
was added to the DIS and DY data sets in the EPS09 analysis 
leading to the nuclear shadowing effect which is weaker 
than that in EPS08, but still stronger than in HKN07
and nDS.

One can also compare the nuclear suppression factor
found from the analysis of the ALICE data to the nuclear gluon 
shadowing factors calculated 
in the leading twist theory of nuclear shadowing~\cite{Frankfurt:2011cs}.
The latter is a dynamical approach   
based on the QCD factorization theorems,
Gribov's theory \cite{gribov} of inelastic shadowing corrections 
in multiple scattering, and the HERA diffractive PDFs~\cite{heradif}.
The comparison is presented in the bottom panel of Fig.~\ref{shad}. We show three 
sets of predictions corresponding to three different sets of the 
gluon PDF in the free proton.
The blue band shows
$R(x,\mu ^2=2.4 \,\,{\rm GeV}^2)$  
calculated with the MNRT07LO nucleon gluon density obtained  
by the Durham--PNPI group~\cite{Martin:2007sb} from the fit to the HERA data
on the coherent photo- and electroproduction of $J/\psi$ mesons on 
the proton.  The uncertainty of the predicted values of 
$R(x,\mu ^2=2.4 \,\,GeV^2)$ (the width of the band)
reflects 
the theoretical uncertainty of the leading twist theory 
associated with the need to model 
the interactions with $N \geq 3$ nucleons of the nuclear target.

The red band shows
$R(x,\mu ^2=2.4 \,\,{\rm GeV}^2)$
calculated using the MSTW08LO nucleon gluon density~\cite{Martin:2009iq}.
The resulting value of $R(x,\mu ^2=2.4 \,\,{\rm GeV}^2)$  is close to the value 
of nuclear suppression  found in the current analysis of the ALICE data
at $x\approx 0.01$, especially if one takes into account
large experimental errors of $S_{Pb}(x)$, and agrees well with the ALICE data point
at $x\approx 0.001$. 

Weaker nuclear gluon shadowing (green band) 
is predicted 
when one uses the CTEQ6L nucleon gluon distribution.

A word of caution is in order here. In the present comparison 
with the leading twist approximation
(LTA) predictions,  we used a particular plausible value of the hard scale $\mu^2$, 
$\mu ^2 =M_{J/\psi}^2/4=2.4$ GeV$^2$.
An increase of $\mu^2$ (within the allowed limits) will result
in weakening of the nuclear gluon shadowing. 
However, one has to keep in mind that a strong increase of 
$\mu ^2$ (small gluon shadowing) is disfavored by the comparison with the ALICE suppression
factors.

In summary,  
our analysis shows that the bulk of the nuclear suppression
found in the ALICE measurements of $J/\psi$ photoproduction in
UPCs of heavy ions at the LHC energies is due to the 
 nuclear gluon shadowing. Moreover, it seems that 
a  reasonable agreement between the measured nuclear suppression factor with
the predictions of nuclear gluon shadowing in the leading twist approximation 
can be considered as evidence of the adequate description 
of this phenomenon in the leading twist framework.

\section{Conclusions}

The analysis of the ALICE measurements of exclusive $J/\psi$ production 
in ultraperipheral PbPb collisions at 2.76 TeV  demonstrates that this data provides 
the experimental evidence for a 
comparatively large nuclear suppression in Pb at small $x \sim 10^{-3}$. 
The found values are in agreement with the nuclear gluon shadowing predicted in the framework 
of leading twist nuclear shadowing and found in the central EPS09LO set of nuclear PDFs.

The significant source of uncertainties comes from experimental errors in 
the measured cross sections both in exclusive $J/\psi$ production in PbPb UPCs 
 and in studies of the $\gamma p\rightarrow J/\psi p$ elementary process. 
 Obviously, more detailed studies of
 $J/\psi$ production in nucleus--nucleus UPCs  at the LHC would be extremely useful 
to put stronger constraints on the nuclear gluon shadowing and gluon distributions in 
nuclei at small $x$.

\section{Acknowledgements}

We would like to thank L.~Frankfurt and H.~Honkanen for useful discussions and H.~Honkanen
for providing the EPS09 predictions used in the middle panel of Fig.~\ref{shad}. 
 This work was supported in part by US DOE Contract Number DE-FG02-93ER40771.

\bibliographystyle{elsarticle-num}

\end{document}